\documentclass[12pt]{iopart}

\usepackage{graphicx}
\usepackage{epstopdf}
\begin{document}

\title{Evolution of superconducting and transport properties in annealed FeTe$_{1-x}$Se$_{x}$ (0.1 $\leq$ \emph{x} $\leq$ 0.4) multiband superconductors}

\author{Yue Sun$^{*}$, Toshihiro Taen, Tatsuhiro Yamada, Yuji Tsuchiya, Sunseng Pyon, Tsuyoshi Tamegai}

\address{Department of Applied Physics, The University of Tokyo, 7-3-1 Hongo, Bunkyo-ku, Tokyo 113-8656, Japan}

\begin{abstract}
We investigated the superconducting and transport properties in FeTe$_{1-x}$Se$_{x}$ (0.1 $\leq$ \emph{x} $\leq$ 0.4) single crystals prepared by O$_2$-annealing. Sharp superconducting transition width observed in magnetization measurement and the small residual resistivity prove the high quality of the crystals. All the crystals manifest large, homogeneous, and isotropic critical current density \emph{J}$_c$ with self-field value over 10$^5$ A/cm$^2$ at 2 K. The large and field-robust critical current densities prove that the superconductivity in FeTe$_{1-x}$Se$_{x}$ (0.1 $\leq$ \emph{x} $\leq$ 0.4) is in bulk nature. The values of anisotropy parameter close to $T_c$ for crystals with different Se doping levels all reside in the range of 2 - 3. Hall coefficients $R_H$ keeps positive and almost constant value at high temperatures, followed by a sudden decreases before reaching \emph{T}$_c$, which indicates that the electron-type charge carriers become dominant at low temperatures. Furthermore, the characteristic temperature for the sudden decrease in $R_H$ gradually increases with  Se doping.
\end{abstract}

\maketitle

\section{Introduction}
Discovered in 2008, iron-based superconductors (IBSs) with superconducting transition temperature \emph{T}$_c$ over 55 K is the second member of the unconventional high temperature superconductors (HTSs) after cuprate superconductors \cite{KamiharaJACS,StewartIBSsreview}. Although IBSs share some similarities with cuprate superconductors like layered structures, very high upper critical fields, and doping phase diagrams, important differences exist between the two families. The cuprates are doped Mott insulators with strong correlation and a single band behavior, while the IBSs show multiband electronic structures \cite{StewartIBSsreview}. Existence of disconnected Fermi surfaces with  electron and hole characters, and spin or orbital fluctuations are supposed to be responsible for the high value of \emph{T}$_c$  in IBSs based on either  \emph{s}$_\pm$ \cite{MazinS} or \emph{s}$_{++}$ scenario \cite{KontaniPRL}. The nesting between electron and hole bands is supposed to be related to the high value of \emph{T}$_c$ in IBSs based on the scenario of \emph{s}$_\pm$ paring. Among the family of IBSs, FeTe$_{1-x}$Se$_{x}$ composed of only Fe(Te,Se) layers has attracted special attention due to its simple crystal structure, which is preferable for probing the mechanism of superconductivity. Band structure calculations and angle-resolved photoemission spectroscopy (ARPES) prove the multiband structure in Fe$_{1+y}$(Te/Se) \cite{SubediPRB,XiaPRL,ChenPRB11}, which is characterized by hole bands around $\Gamma$ point and electron bands around M point, similar to iron pnictides. On the other hand, its less toxic nature makes FeTe$_{1-x}$Se$_{x}$ a more suitable candidate for applications among the family of IBSs.

Although much attention has been paid to this compound, some crucially fundamental properties are still controversial or unrevealed. In optimally-doped FeTe$_{1-x}$Se$_{x}$, ARPES shows an isotropic superconducting gap \cite{MiaoPRB}, while anisotropic or two-gap features were suggested by angle-resolved specific heat measurements \cite{ZengNatCommun}, muon spin rotation measurements \cite{BendelePRB, BiswasPRB}, and optical conductivity measurements \cite{PimenovNJP}. Recently, London penetration depth measurement down to 50 mK indicates a possible nodal gap \cite{DiaconuPRB}. On the other hand, although Liu \emph{et al}. \cite{LiuNatMat} reported that bulk superconductivity resides only in the region of Se level higher than 29\%, it was observed in the Se doping level between 10 and 50\% by Noji \emph{et al}. \cite{NojiJPSJPhaseD}. As for the Hall coefficient, conflicting low-temperature behaviors have been observed in crystals with nominally the same amount of Se \cite{LiuPRB,LiuSUST,TsukadaPRB}. Even in the case of resistivity, both the metallic and nonmetallic behaviors were reported, and the absolute value just above \emph{T}$_c$ has a spread from 200 to 1500 $\mu$$\Omega$cm \cite{NojiJPSJPhaseD, LiuPRB, LiuSUST, TsukadaPRB}. On the other hand, the evolution of some essential parameters, like the critical current density \emph{J}$_c$, upper critical field \emph{H}$_{c2}$, and anisotropy parameter $\gamma$ with Se-doping level are still unrevealed. All these controversies and unrevealed issues are believed to come from the sample-dependent Fe nonstoichiometries \cite{BaoWeiPRL, BendelePRB2010}, which originate from the partial occupation of the second Fe site (excess Fe site) in the Te/Se layer. The excess Fe with valence near Fe$^+$ will provide electron doping into the 11 system \cite{ZhangPRB}. The excess Fe is also strongly magnetic, which produces local moments that interact with the adjacent Fe layers \cite{ZhangPRB}. The magnetic moment from excess Fe will act as a pair breaker and also localize the charge carriers \cite{LiuPRB}.

Previous reports have proved that post annealing is effective to remove the excess Fe in FeTe$_{1-x}$Se$_{x}$ single crystal \cite{TaenPRB,DongPRB,KawasakiSSC,HuSUST,RodriguezJACs,SunSUST,Sunjpsj,SunJPSJshort,ZhouSUST}. Our previous STM result directly proved that the excess Fe was totally removed after annealing with appropriate amount of O$_2$, and the large value of normalized specific heat jump at $T_c$ manifested the high-quality of the annealed crystals \cite{SunSciRep}. In this paper, we benefit from those technique to perform careful study of the evolution of superconducting and tranport properties in high-quality FeTe$_{1-x}$Se$_{x}$ (0.1 $\leq$ \emph{x} $\leq$ 0.4) single crystals. The evolution of \emph{J}$_c$, \emph{H}$_{c2}$, and $\gamma$ with Se doping in FeTe$_{1-x}$Se$_{x}$ were reported for the first time in high-quality single crystals. Besides, the influence of Se doping to the transport property is also reported for the first time on crystals without the influence of excess Fe.

\section{Experimental details}
Single crystals with nominal compositions FeTe$_{1-x}$Se$_{x}$ (\emph{x} = 0.1, 0.2, 0.3, 0.4) were grown by the self-flux method \cite{SunSUST}. All the crystals show mirror-like surface with the size reaching centimeter-scale as shown in Fig. 1 (a). The actual Se doping level is $\sim$ 0.09, 0.20, 0.33 and 0.43, respectively, detected by the Inductively-coupled plasma (ICP) atomic emission spectroscopy and energy dispersive x-ray spectroscopy (EDX) measurements. Fig. 1(b) shows the SEM photo of the as-grown FeTe$_{0.6}$Se$_{0.4}$ single crystal. Compositional mappings by EDX in the selected rectangular region of Fig. 1(b) are shown in Fig. 1(c) and (d), which prove that Te and Se are almost homogeneously distributed in the crystal. Actually, the doping composition and homogeneity change little after annealing. In addition, compositional measurements also manifest that the obtained as-grown single crystals usually contain $\sim$14\% excess Fe. To remove the excess Fe, the as-grown crystals were further annealed in the O$_2$ atmosphere. Details of the crystal growth and annealing  procedure have been reported in our previous publications \cite{SunSUST,SunSciRep}. Magnetization measurements were performed using a commercial SQUID magnetometer (MPMS-XL5, Quantum Design). Magneto-optical (MO) images were obtained by using the local field-dependent Faraday effect in the in-plane magnetized garnet indicator film employing a differential method \cite{MOexperimentalnature, MOexperimentalPRB}. Longitudinal and transverse (Hall) resistivities were measured by six-lead method with a Physical Property Measurement System (PPMS, Quantum Design) at temperatures down to 2 K and magnetic fields up to 9 T. In order to decrease the contact resistance, we sputtered gold on the contact pads just after the cleavage, then gold wires were attached on the contacts with silver paste.
\begin{figure}\center

　　\includegraphics[width=8cm]{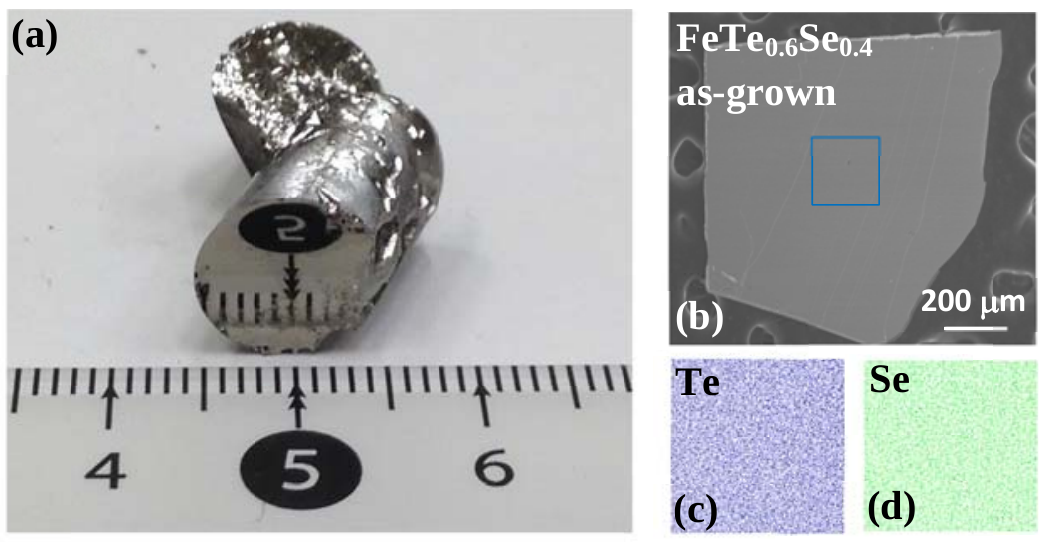}\\
　　\caption{(a) Photograph of the as-grown FeTe$_{0.6}$Se$_{0.4}$ single crystal. (b) SEM photo of a cleaved as-grown FeTe$_{0.6}$Se$_{0.4}$ single crystal. (c) and (d) are the Te and Se distribution probed by EDX analyses in the selected rectangular region of (b) .  }\label{}
\end{figure}

\section{Results and discussion}
\begin{figure}\center

　　\includegraphics[width=8cm]{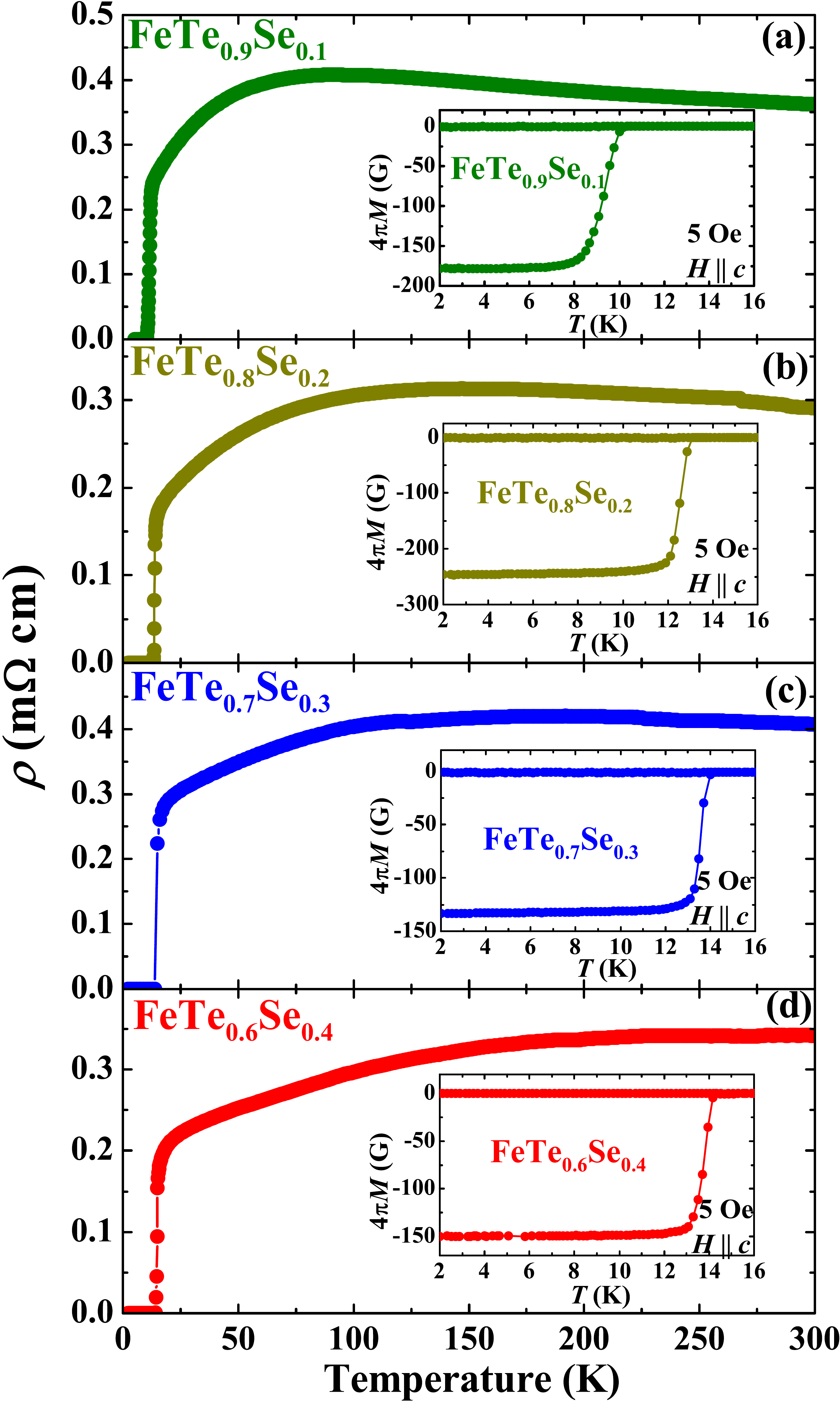}\\
　　\caption{Temperature dependence of resistivity of annealed (a) FeTe$_{0.9}$Se$_{0.1}$, (b) FeTe$_{0.8}$Se$_{0.2}$, (c) FeTe$_{0.7}$Se$_{0.3}$, and (d) FeTe$_{0.6}$Se$_{0.4}$  single crystals. Insets show the temperature dependences of ZFC and FZ magnetizations at 5 Oe parallel to $c$-axis for the corresponding crystals.}\label{}
\end{figure}

Figure 2(a) - (d) show the temperature dependence of in-plane resistivity for the annealed FeTe$_{0.9}$Se$_{0.1}$, FeTe$_{0.8}$Se$_{0.2}$, FeTe$_{0.7}$Se$_{0.3}$, and FeTe$_{0.6}$Se$_{0.4}$  single crystals, respectively. It is obvious that resistivity maintains a nearly constant value or slightly increase with decreasing temperature at high temperatures, followed by a metallic behavior (d$\rho$/d$T$ $>$ 0) below 100 $\sim$ 150 K. Actually, some previously reported resistive results on FeTe$_{1-x}$Se$_{x}$ show semiconducting behavior before superconducting transition temperature \cite{LiuPRB, LiuSUST, TsukadaPRB}, which is interpreted by the effect of excess Fe \cite{LiuPRB, SunPRB}. Density functional study shows that the excess Fe in the interstitial sites is magnetic and interacts with magnetism of Fe in Fe planes \cite{ZhangPRB}. The magnetic moment from excess Fe will localize the charge carriers, which causes the semiconducting resistive behavior. The metallic resistive behavior observed in all the four annealed crystals show that the effect of excess Fe was minimized after O$_2$ annealing. The resistivity just above \emph{T}$_c$ is $\sim$ 200 $\mu$$\Omega$cm, which is the smallest among those reported in Fe(Te,Se) crystals and films \cite{LiuNatMat,LiuPRB,LiuSUST,TaenPRB,TsukadaPRB,SalesPRB}, manifesting the high quality of the annealed crystals.

To further confirm the superconducting transition temperature, \emph{T}$_c$,  as well as the sample quality, temperature dependences of zero-field-cooled (ZFC) and field-cooled (FC) magnetizations were measured and shown in the insets of Fig. 2, from which the \emph{T}$_c$ can be determined as 10.1 K, 12.9 K, 14.0 K and 14.3 K for FeTe$_{0.9}$Se$_{0.1}$, FeTe$_{0.8}$Se$_{0.2}$, FeTe$_{0.7}$Se$_{0.3}$, and FeTe$_{0.6}$Se$_{0.4}$, respectively. Besides, all the crystals show sharp superconducting transitions, confirming the high quality of our annealed FeTe$_{1-x}$Se$_{x}$ single crystals.
\begin{figure}\center

　　\includegraphics[width=15cm]{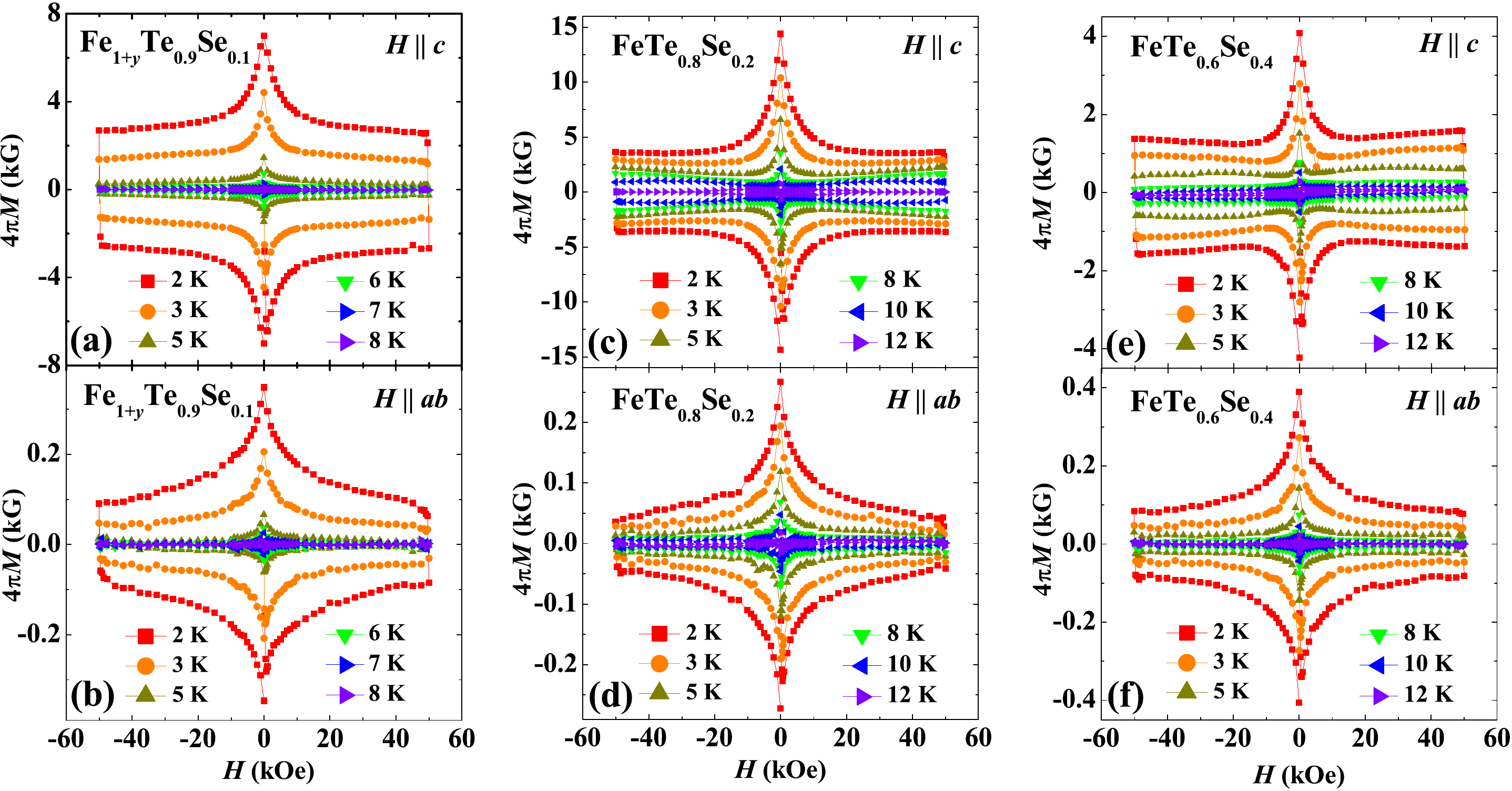}\\
　　\caption{Magnetic hysteresis loops (MHLs) at several temperatures measured for \emph{H} $\|$ \emph{c} and \emph{H} $\|$ \emph{ab} in annealed (a)-(b) FeTe$_{0.9}$Se$_{0.1}$, (c)-(d) FeTe$_{0.8}$Se$_{0.2}$, and (e)-(f) FeTe$_{0.6}$Se$_{0.4}$.}\label{}
\end{figure}

To further probe the superconducting and magnetic properties of  FeTe$_{1-x}$Se$_{x}$ single crystals, magnetic hysteresis loops (MHLs) at several temperatures were measured for \emph{H} $\|$ \emph{c} and \emph{H} $\|$ \emph{ab}. Typical results for FeTe$_{0.9}$Se$_{0.1}$, FeTe$_{0.8}$Se$_{0.2}$, and FeTe$_{0.6}$Se$_{0.4}$ are shown in Figure 3(a)-(b), (c)-(d), (e)-(f), respectively. In the case of \emph{H} $\|$ \emph{c}, a second magnetization peak (SMP), also known as the fish-tail effect (FE), can be witnessed, although not very prominent because the peak position is beyond the maximum applied field at low temperatures. On the other hand, for \emph{H} $\|$ \emph{ab}, the scenario is quite different compared with the situation for \emph{H} $\|$ \emph{c}. Here, the SMP is not observed within our measured range of magnetic field. This difference can be seen more clearly in magnetic field dependent critical current density in Figure 4 obtained by using the extended Bean model \cite{Beanmodel}:
\begin{equation}
\label{eq.1}
J_c=20\frac{\Delta M}{a(1-a/3b)},
\end{equation}
where $\Delta$\emph{M} is \emph{M}$_{down}$ - \emph{M}$_{up}$, \emph{M}$_{up}$ [emu/cm$^3$] and \emph{M}$_{down}$ [emu/cm$^3$] are the magnetization when sweeping fields up and down, respectively, \emph{a} [cm] and \emph{b} [cm] are sample widths (\emph{a} $<$ \emph{b}). Actually, in tetragonal two-dimensional systems, there are three kinds of critical current density, \emph{J}$_c^{x,y}$, where \emph{x} and \emph{y} refer to the directions of current and magnetic field, respectively. For \emph{H} $\|$ \emph{c}, irreversible magnetization is determined solely by \emph{J}$_c^{ab,c}$. This means that \emph{J}$_c^{ab,c}$ (= \emph{J}$_c^{H||c}$) can be easily evaluated from the measured magnetization using the extended Bean model. On the other hand, in the case of \emph{H} $\|$ \emph{ab}, both \emph{J}$_c^{ab,ab}$ and \emph{J}$_c^{c,ab}$ contribute to the measured magnetization. Here we simply assume that \emph{J}$_c^{ab,ab}$ is equal to \emph{J}$_c^{c,ab}$ \cite{SunAPRE}, and obtain the weighted average \emph{H} $\|$ \emph{ab} using eq.(1). Based on the discussion above, self-field critical current density \emph{J}$_c^{H||c}$ and \emph{J}$_c^{H||ab}$ at 2 K are estimated as 4 $\times$ 10$^5$ A/cm$^2$ and 3.5 $\times$ 10$^5$ A/cm$^2$ for FeTe$_{0.9}$Se$_{0.1}$, 5.8 $\times$ 10$^5$ A/cm$^2$ and 4.1 $\times$ 10$^5$ A/cm$^2$ for FeTe$_{0.8}$Se$_{0.2}$, 3 $\times$ 10$^5$ A/cm$^2$ and 2.5 $\times$ 10$^5$ A/cm$^2$ for FeTe$_{0.6}$Se$_{0.4}$, respectively. Thus, the critical current densities are large, almost isotropic, and not sensitive to the Se doping level. The large critical current densities also prove that the superconductivity observed in FeTe$_{1-x}$Se$_{x}$ (0.1 $\leq$ \emph{x} $\leq$ 0.4) is in bulk nature. Furthermore, \emph{J}$_c$'s are also robust against the applied field at low temperatures.
\begin{figure}\center

　　\includegraphics[width=15cm]{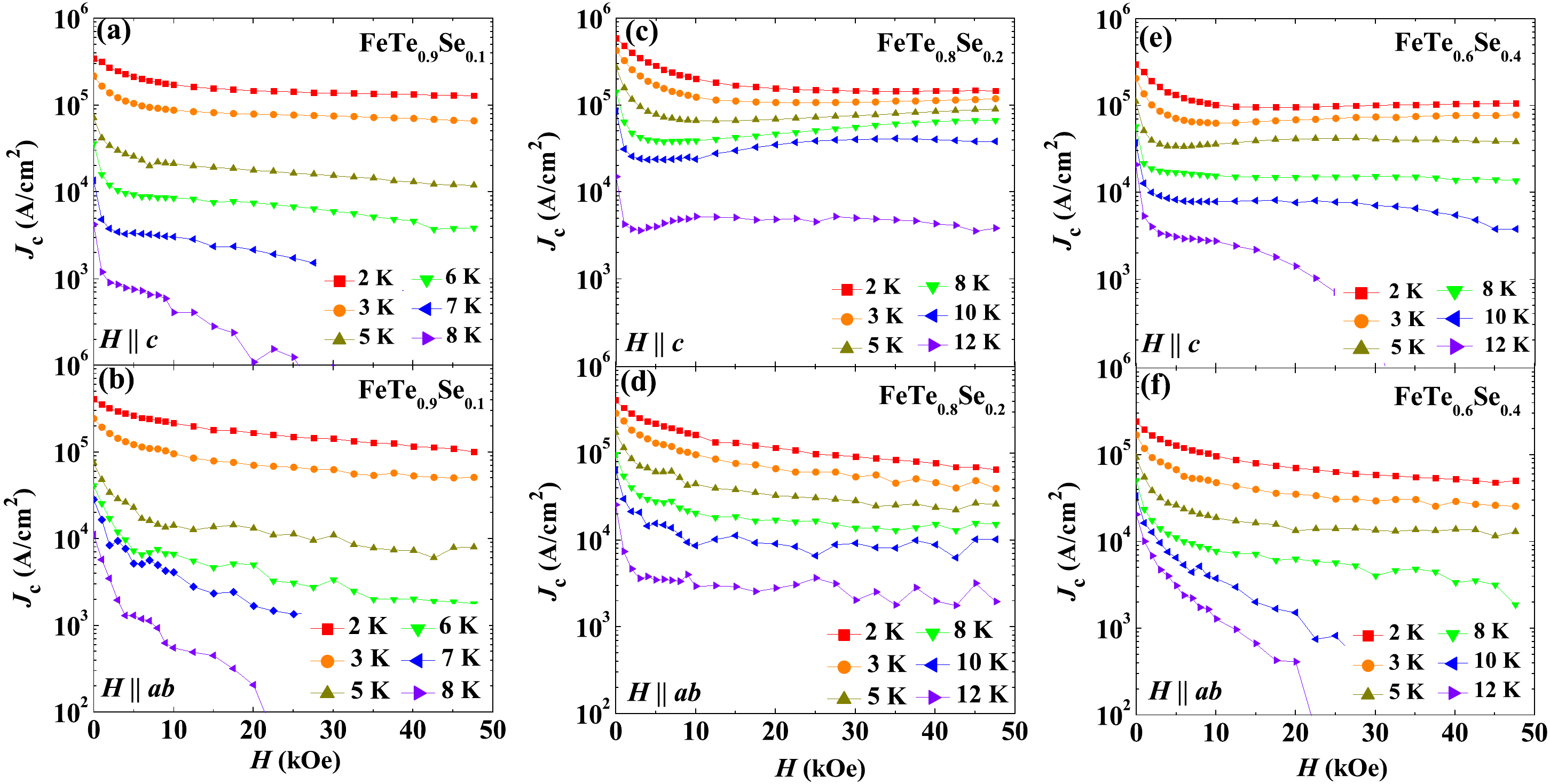}\\
　　\caption{Critical current densities \emph{J}$_c$ at several temperatures measured for \emph{H} $\|$ \emph{c} and \emph{H} $\|$ \emph{ab} in annealed (a)-(b) FeTe$_{0.9}$Se$_{0.1}$, (c)-(d) FeTe$_{0.8}$Se$_{0.2}$, and (e)-(f) FeTe$_{0.6}$Se$_{0.4}$. }\label{}
\end{figure}

To further check the distribution of \emph{J}$_c$ in the crystals, we took MO images on the four crystals in the remanent state, which are prepared by applying magnetic field, large enough to totally penetrate the sample, along \emph{c}-axis, and removing it after zero-field cooling. Typical MO images at selected temperatures for FeTe$_{0.9}$Se$_{0.1}$, FeTe$_{0.8}$Se$_{0.2}$, FeTe$_{0.7}$Se$_{0.3}$, and FeTe$_{0.6}$Se$_{0.4}$  single crystals are shown in Figure 5(a)-(d), respectively. The MO images manifest typical roof-top patterns, indicating a nearly uniform current flow in the crystal. Besides, typical current discontinuity lines (so-called \emph{d}-line), which cannot be crossed by vortices, can be observed and marked by the dashed lines. The angle $\theta$ of the \emph{d}-line for all the four crystals is $\sim$ 45$^\circ$, indicating that the critical current density within the \emph{ab} plane is isotropic, consistent with the four-fold symmetry of the superconducting plane. The above results proved that the superconductivity observed in FeTe$_{1-x}$Se$_{x}$ single crystals with the Se doping level ranging from 0.1 to 0.4 is in bulk nature, and the critical current densities are large, isotropic, and almost homogeneously distributed in the crystals. In addition, the value of \emph{J}$_c$ is not sensitive to the Se doping level, which is also an advantage to application.
\begin{figure}\center
　　\includegraphics[width=15cm]{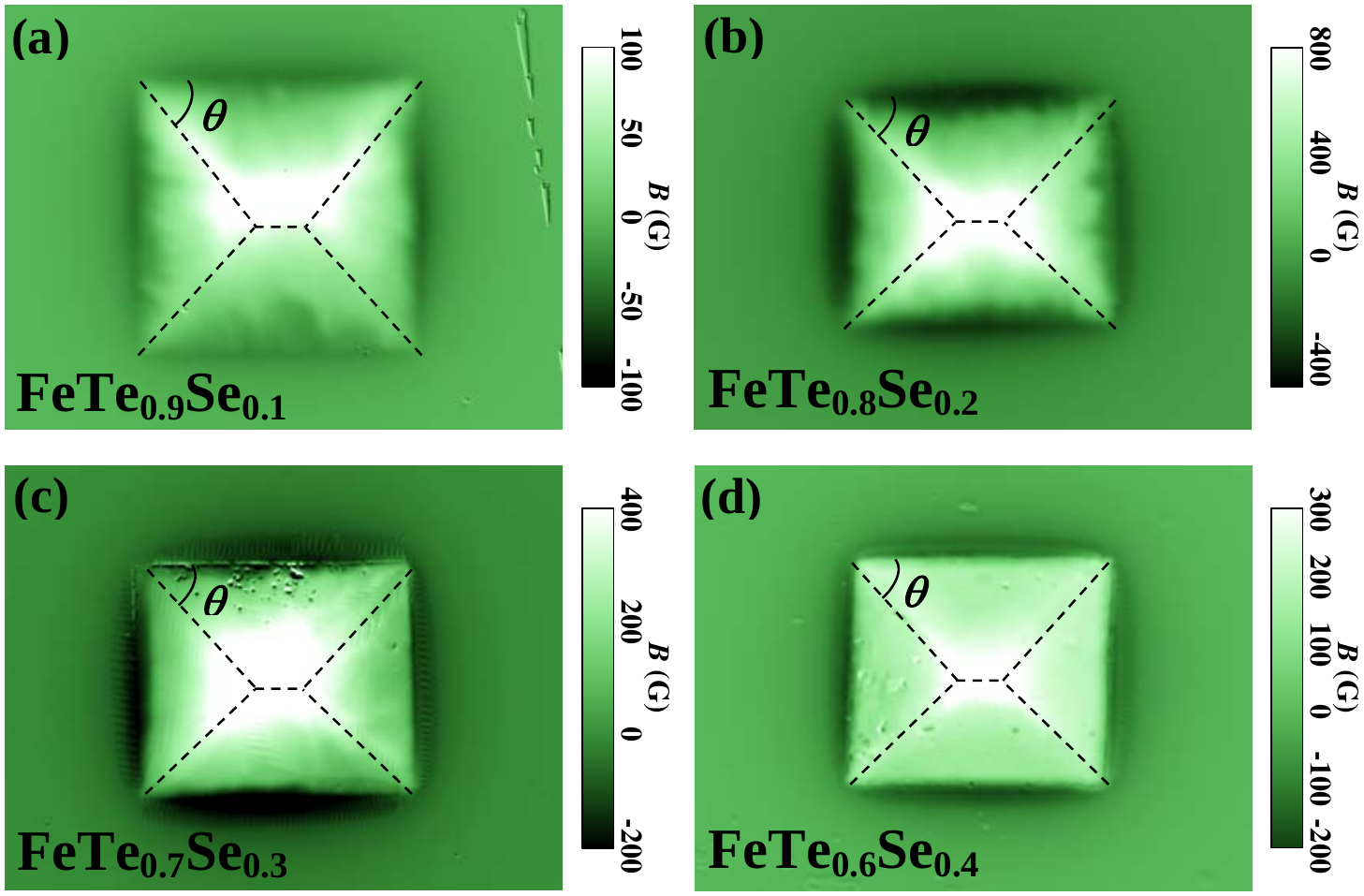}\\
　　\caption{Magneto-optical images in the remanent state for annealed single crystals (a) FeTe$_{0.9}$Se$_{0.1}$ at 7 K, (b) FeTe$_{0.8}$Se$_{0.2}$ at 11 K, (c) FeTe$_{0.7}$Se$_{0.3}$ at 9 K, and (d) FeTe$_{0.6}$Se$_{0.4}$ at 8 K after applying 800 Oe along the \emph{c}-axis.}\label{}
\end{figure}

Figure 6(b) and (c), (e) and (f), (h) and (i) show the temperature dependence of resistivities with applied field \emph{H} $\|$ \emph{c} and \emph{H} $\|$ \emph{ab} from 0 to 9 T for FeTe$_{0.9}$Se$_{0.1}$, FeTe$_{0.8}$Se$_{0.2}$ and FeTe$_{0.6}$Se$_{0.4}$ single crystals, respectively. Obviously, for crystals with different Se doping levels, when \emph{H} $\|$ \emph{c}, the resistive transition shifts to lower temperatures with increasing field, accompanied by a slight increase in the transition width, while this broadening is almost negligible for \emph{H} $\|$ \emph{ab}. The upper critical fields \emph{H}$_{c2}$ for \emph{H} $\|$ \emph{c} and \emph{H} $\|$ \emph{ab} defined by the midpoint of the superconducting transition are plotted in Fig. 6(a), (d), and (g) for FeTe$_{0.9}$Se$_{0.1}$, FeTe$_{0.8}$Se$_{0.2}$, and FeTe$_{0.6}$Se$_{0.4}$, respectively. Obviously, the \emph{H}$_{c2}$(\emph{T}) curves for \emph{H} $\|$ \emph{ab} show a clear decrease of slope with decreasing temperature, whereas for \emph{H} $\|$ \emph{c} the curves remain almost linear. The early saturation of the slope in \emph{H}$_{c2}^{ab}$ reflects Pauli paramagnetic limit \cite{PaulilimiteAPL,PaulilimitePRL}. Similar saturating behavior is also reported in other iron based superconductors \cite{GurevichRepProg}. According to the Werthamer-Helfand-Hohenberg theory \cite{WHH}, \emph{H}$_{c2}$ limited by the orbital depairing in the dirty limit is given by \emph{H}$_{c2}$(0) = -0.693\emph{T}$_c$\emph{d}\emph{H}$_{c2}$/\emph{dT}$\mid$$_{T=T_c}$. Thus, the \emph{H}$_{c2}^c$(0) and \emph{H}$_{c2}^{ab}$(0) can be estimated with the slopes of the linear part below 4 T as $\sim$43.5 T and $\sim$85.1 T for FeTe$_{0.9}$Se$_{0.1}$, $\sim$64.2 T and $\sim$115.0 T for FeTe$_{0.8}$Se$_{0.2}$ and $\sim$64.0 T and $\sim$132.4 T for FeTe$_{0.6}$Se$_{0.4}$. Because of the Pauli paramagnetic limit, the actual upper critical field at low temperatures, especially for $H$ $\|$ $ab$  should be smaller than the estimated values above.

Insets of Fig. 6(a), (d), and (g) show the temperature dependence of anisotropy $\gamma$ = \emph{H}$_{c2}^{ab}$/\emph{H}$_{c2}^c$ for FeTe$_{0.9}$Se$_{0.1}$, FeTe$_{0.8}$Se$_{0.2}$, and FeTe$_{0.6}$Se$_{0.4}$, respectively. The values of $\gamma$ close to $T_c$ for all the three crystals reside in the range of 2 - 3, similar to the case of 122-type IBSs \cite{GurevichRepProg}. Besides, the values of $\gamma$ show slightly decreasing with decreasing temperature, which reflects the saturating behavior for \emph{H} $\|$ \emph{ab} caused by Pauli paramagnetic effect. The anisotropy of the upper critical field presents important information about the anisotropy of the electrical resistivity. In the temperature range close to the zero-field \emph{T}$_c$, the anisotropy of \emph{H}$_{c2}$ for an \emph{s}-wave superconductor can be written as $\gamma$ $\sim$ \emph{v}$_a$/\emph{v}$_c$, where \emph{v}$_a$ and \emph{v}$_c$ are Fermi velocities for \emph{a} and \emph{c} directions, respectively \cite{GokovJETP}. This approximation was verified semi-quantitatively in iron-based superconductors \cite{TanatarPRB2009,LiuPRB2014Hc2}. Thus, similar $\gamma$ values obtained in FeTe$_{1-x}$Se$_{x}$ (0.1 $\leq$ $x$ $\leq$ 0.4) indicates that the topology of the Fermi surface changes little in this range.
\begin{figure}\center

　　\includegraphics[width=15cm]{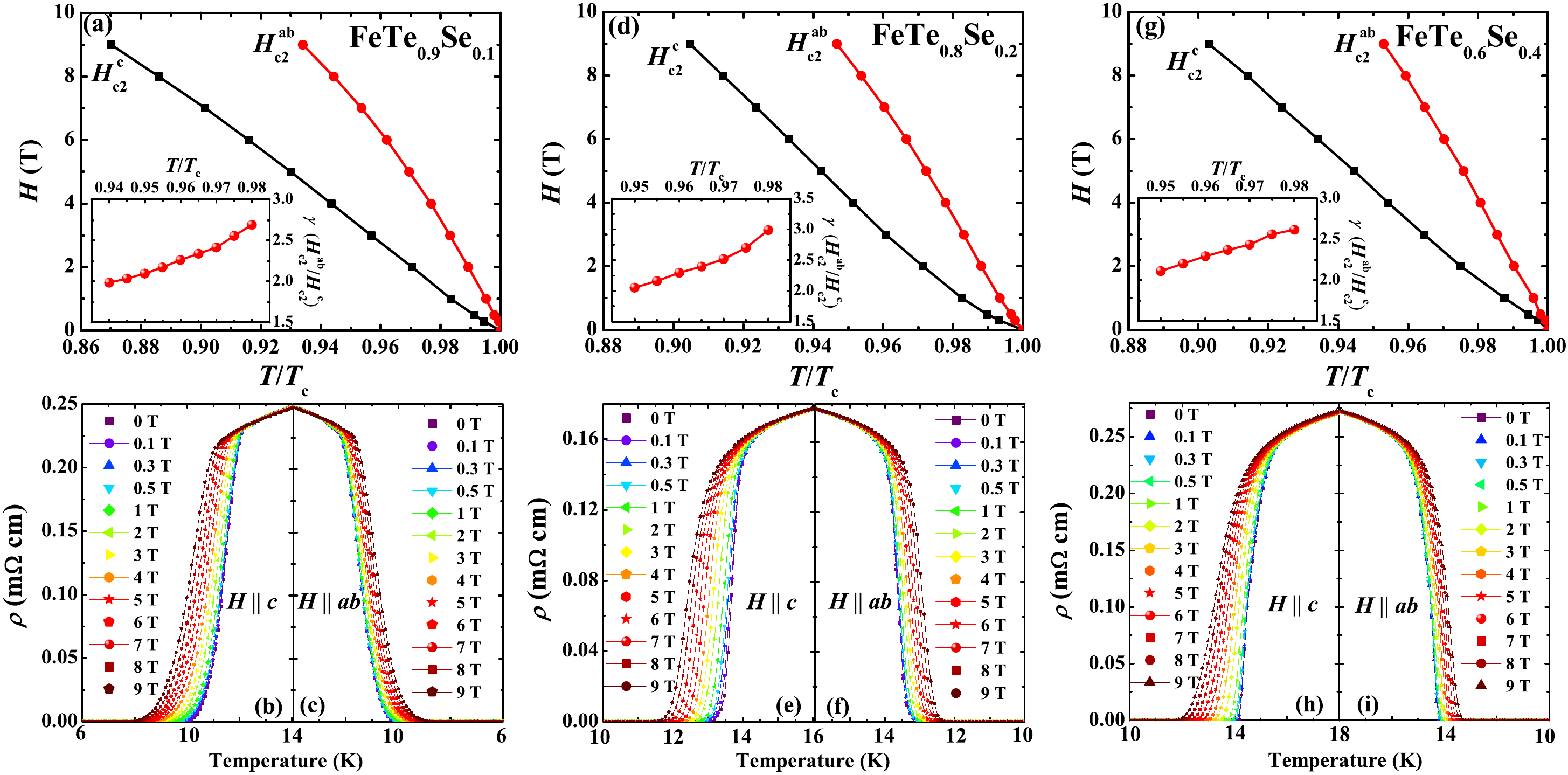}\\
　　\caption{Temperature dependence of resistivities with applied field \emph{H} $\|$ \emph{c} and \emph{H} $\|$ \emph{ab} from 0 to 9 T for (b)-(c) FeTe$_{0.9}$Se$_{0.1}$, (e)-(f) FeTe$_{0.8}$Se$_{0.2}$, and (h)-(i) FeTe$_{0.6}$Se$_{0.4}$ single crystals. Temperature dependence of upper critical fields is shown in (a) FeTe$_{0.9}$Se$_{0.1}$, (d) FeTe$_{0.8}$Se$_{0.2}$, and (g) FeTe$_{0.6}$Se$_{0.4}$. Insets of (a), (d), and (g) show the temperature dependence of anisotropy parameter $\gamma$ for FeTe$_{0.9}$Se$_{0.1}$, FeTe$_{0.8}$Se$_{0.2}$, and FeTe$_{0.6}$Se$_{0.4}$, respectively}\label{}
\end{figure}

All the above results confirm the high quality of our annealed FeTe$_{1-x}$Se$_{x}$ single crystals. Now we turn to the evolution of the transport properties with the Se doping in those high-quality crystals. Figure 7 shows the temperature dependence of the Hall coefficient for FeTe$_{1-x}$Se$_{x}$ (\emph{x} = 0.1, 0.2, 0.3, 0.4) single crystals. To clearly show the influence of sample quality to the transport properties, results of Hall coefficients for the four crystals before annealing were also plotted in the figure with open symbols for comparison. Obviously, \emph{R}$_H$ is almost temperature independent above 100 K, and keeps a constant value $\sim$1 $\times$ 10$^{-9}$ m$^3$/C for all the crystals before and after annealing. When temperature is decreased below 100 K, \emph{R}$_H$ starts to increase. In the as-grown crystals, \emph{R}$_H$  gradually increases with decreasing temperature showing an obvious divergence at low temperatures, except for FeTe$_{0.9}$Se$_{0.1}$. Such divergent behavior is attributed to the effect of excess Fe, which is magnetic and interacts with the plane Fe magnetism proved by the density functional study \cite{ZhangPRB}. The magnetic moment produced by excess Fe will localize the charge carriers, and cause the upturn in \emph{R}$_H$ \cite{LiuPRB, SunPRB}. The upturn behavior becomes flattened in the as-grown FeTe$_{0.9}$Se$_{0.1}$ single crystal, which may come from the remaining antiferromagnetism according to the phase diagram \cite{DongPRB}. After removing the excess Fe by O$_2$ annealing, the divergent increase was totally suppressed in the annealed crystals. The comparison of transport properties in crystals before and after annealing proves that the previously reported upturn behavior of \emph{R}$_H$ in FeTe$_{1-x}$Se$_{x}$ \cite{LiuPRB, LiuSUST, TsukadaPRB} mainly comes from the sample quality, i.e., existence of excess Fe.

\begin{figure}\center

　　\includegraphics[width=8cm]{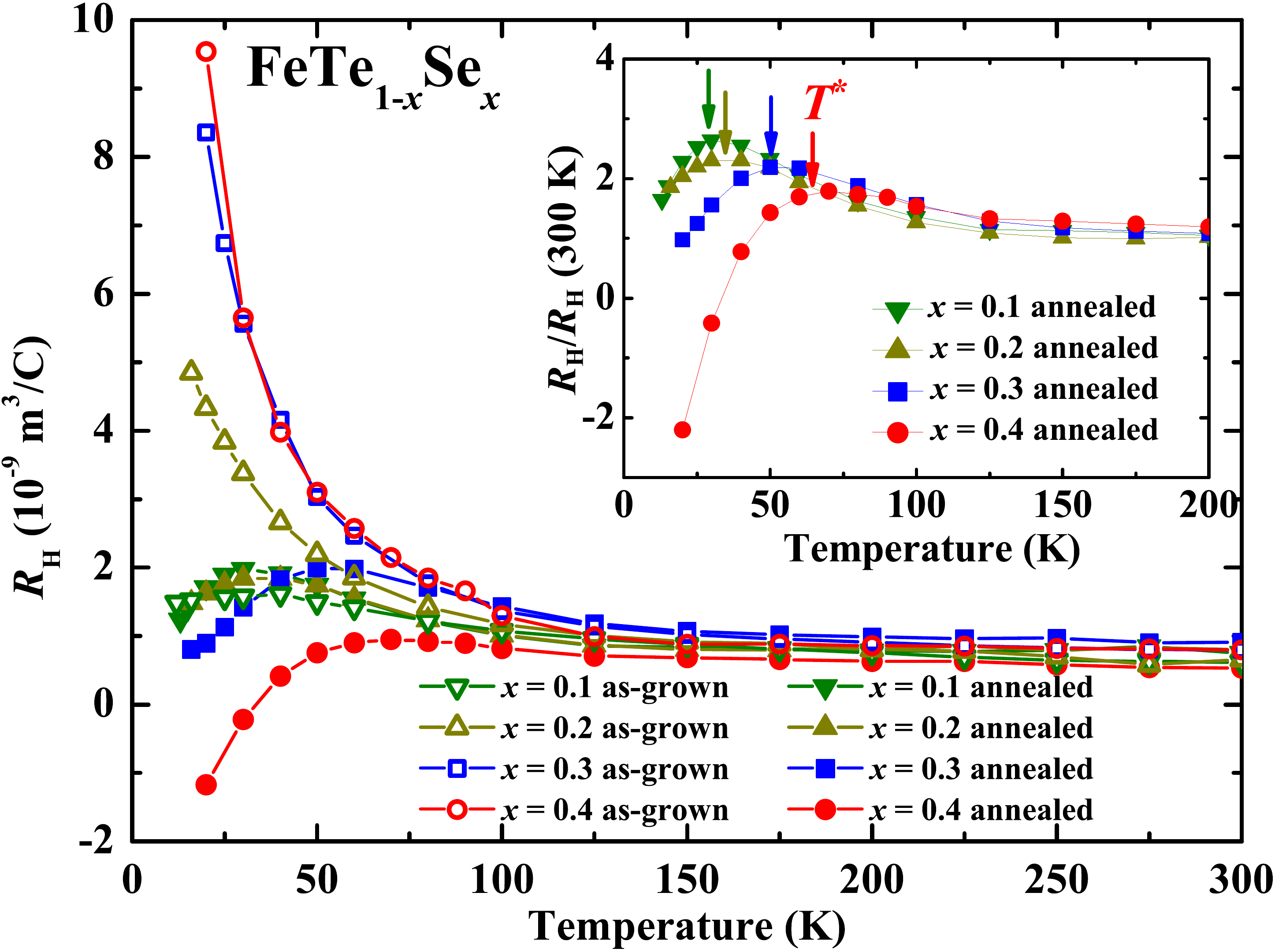}\\
　　\caption{Temperature dependence of Hall coefficients for the as-grown (open symbols) and annealed (solid symbols) FeTe$_{1-x}$Se$_{x}$ (0.1 $\leq$ \emph{x} $\leq$ 0.4) single crystals. Inset shows the Hall coefficients scaled by the values at 300 K in the temperature region of 0 - 200 K.}\label{}
\end{figure}

To clearly show the evolution of transport property with Se doping in the annealed FeTe$_{1-x}$Se$_{x}$ single crystals, we scaled the Hall coefficients by the values at 300 K, and plotted in the inset of Fig. 7. In the annealed crystals, \emph{R}$_H$ keeps nearly temperature independent behavior at high temperatures, followed by a slight increase below 100 K, then suddenly decreases before reaching \emph{T}$_c$. The value even changes sign from positive to negative in FeTe$_{0.6}$Se$_{0.4}$. The strongly temperature-dependent $R_H$ manifests the multiband nature of FeTe$_{1-x}$Se$_{x}$. Actually, multiband structure has already been proven to be a common property in IBSs. According to the band structure calculations and  ARPES results, at least four bands originated from Fe 3\emph{d} orbitals cross the Fermi level \cite{SubediPRB,ChenPRB11}. Two of them contribute hole-type charge carriers, and the other two contribute electron-type charge carriers. In the annealed crystals, Hall coefficient keeps an almost constant positive value at high temperatures, which means the hole-type carrier is dominant. The values of charge carrier densities are not largely influenced by Se doping, which is consistent with the isovalent doping of Se. When temperature is decreased below 100 K, the slight increase in \emph{R}$_H$ may come from the mobility change of the hole-type carriers. Before reaching \emph{T}$_c$, \emph{R}$_H$ suddenly decreases, even changes sign to negative for FeTe$_{0.6}$Se$_{0.4}$, which indicates that the electron-type charge carriers become more dominant. We define the characteristic temperature at which \emph{R}$_H$ begins to decrease as $T^*$, and indicated by arrows in the figure. Here, we should point out that the intrinsic anomaly may begin at higher temperatures, and becomes distinct only when temperature is decreased to $T^*$. $T^*$ gradually increases from $\sim$28 K in FeTe$_{0.9}$Se$_{0.1}$ to 65 K in FeTe$_{0.6}$Se$_{0.4}$, indicating that it is related to the Se doping. Actually, our previous report on FeTe$_{0.6}$Se$_{0.4}$ revealed that a large and linear magnetoresistance emerges when temperature decreases below $T^*$, which indicates the possible existence of electron-type Dirac-cone state \cite{SunPRB}. Although the weight of the Dirac cone state is small, it can dominate the transport properties because of its extremely high mobility. Thus, the observed evolution of transport properties in FeTe$_{1-x}$Se$_{x}$ may be originated from its intrinsic multiband effect. On the other hand, such strong temperature and doping dependence of $R_H$ can be also explained in terms of anisotropic charge carrier scattering due to interband antiferromagnetic fluctuations similar to the case of BaFe$_{2-x}$Ru$_x$As$_2$ \cite{EomPRB}. More efforts should be done in future to solve this issue.

\section{Conclusions}
In summary, we have carefully studied the superconducting and transport properties in high-quality FeTe$_{1-x}$Se$_{x}$ (\emph{x} = 0.1, 0.2, 0.3, 0.4) single crystals prepared by O$_2$-annealing. Sharp superconducting transition widths observed in magnetization measurement and small residual resistivities manifest the high quality of the crystals. Large, homogeneous, and almost isotropic critical current densities were obtained in all the annealed crystals with Se doping level ranging from 10\% to 40\%, which proves that the superconductivity in FeTe$_{1-x}$Se$_{x}$ (0.1 $\leq$ \emph{x} $\leq$ 0.4) is in bulk nature. The anisotropy parameter $\gamma$ for all the crystals resides in the range of 2 - 3 obtained from resistivity transition measurements under magnetic field. The values of \emph{J}$_c$ and $\gamma$ are not sensitive to the Se doping, which is an advantage for applications. Temperature dependence of the Hall coefficient keeps positive and almost constant value at high temperatures, followed by a sudden decrease, or even a sign change in FeTe$_{0.6}$Se$_{0.4}$, before reaching \emph{T}$_c$, which indicates that the electron-type charge carriers become dominant at low temperatures. The characteristic temperature for the sudden decrease in $R_H$ gradually increases with  Se doping. All these results in O$_2$-annealed FeTe$_{1-x}$Se$_{x}$ single crystals demonstrate the importance of the annealing process to access the intrinsic properties of this system and unveil its multi-band nature.

\ack
Y.S. gratefully appreciates the support from Japan Society for the Promotion of Science. ICP analyses were performed at Chemical Analysis Section in Materials Analysis Station of NIMS.
\\
\\$^{*}$sunyue.seu@gmail.com
\section*{References}
\bibliographystyle{iopart-num}
\bibliography{references}

\end{document}